\documentclass[letterpaper,twocolumn,10pt]{article}
\usepackage{usenix}

\usepackage{authblk}

\usepackage{tikz}
\usepackage{amsmath}

\usepackage{filecontents}

\usepackage{algorithm}
\usepackage{algpseudocode}

\usepackage{subfig}

\usepackage{makecell}
\usepackage{multirow}
\usepackage{booktabs}

\usepackage{xurl}

\usepackage[available, functional]{usenixbadges}

\begin{document}
\pagenumbering{gobble}
\date{}

\title{\Large \bf Boosting File Systems Elegantly:\\
  A Transparent NVM Write-ahead Log for Disk File Systems}

\author[]{Guoyu Wang}
\author[]{Xilong Che}
\author[]{Haoyang Wei}
\author[]{Shuo Chen}
\author[]{Puyi He}
\author[]{Juncheng Hu}

\affil[]{Jilin University}

\maketitle

\begin{abstract}

We propose NVLog, an NVM-based write-ahead log for disk file systems, designed to transparently harness the high performance of NVM within the legacy storage stack. 
NVLog provides on-demand byte-granularity sync absorption, reserving the fast DRAM path for asynchronous operations, meanwhile occupying NVM space only temporarily. To accomplish this, we designed a highly efficient log structure, developed mechanisms to address heterogeneous crash consistency, optimized for small writes, and implemented robust crash recovery and garbage collection methods.
Compared to previous solutions, NVLog is lighter, more stable, and delivers higher performance, all while leveraging the mature kernel software stack and avoiding data migration overhead. Experimental results demonstrate that NVLog can accelerate disk file systems by up to 15.09x and outperform NOVA and SPFS in various scenarios by up to 3.72x and 324.11x, respectively.

\end{abstract}

\section{Introduction}

Non-volatile memory (NVM), with its persistent and byte-addressable characteristics, is emerging as a new tier in the memory hierarchy, poised to accelerate lower storage devices while ensuring data consistency. 
However, the unique properties of NVM, combined with traditional DRAM and disk, create a complex heterogeneous system. The complexity presents both opportunities and challenges for software design.

To leverage NVM, various NVM-specific file systems have been proposed. Among them, NVM-specialized kernel file systems \cite{DAX, dulloorSystemSoftwarePersistent2014, xuNOVALogstructuredFile2016} are the most popular due to their compatibility with the existing kernel software stack, allowing existing applications to transition to NVM file systems with minimal cost. Unlike traditional block-based file systems, these file systems prioritize the byte-addressable and persistent nature of NVM and achieve higher performance.
However, the significantly lower capacity of NVM compared to block storage devices limits the widespread adoption of NVM-specialized file systems. Additionally, many well-known and widely accepted NVM file systems, such as NOVA, utilize direct access (DAX) to reduce the need for copying data from DRAM to NVM. As we will show in Figure \ref{fig:speed}, the relatively slower speed of NVM can make these file systems less efficient than page cache in many scenarios. 

There are also cross-media file systems \cite{kwonStrataCrossMedia2017, zhengZigguratTieredFile2019} designed to exploit the heterogeneous performance and capacity characteristics of different storage tiers, typically including DRAM, NVM, and block storage. However, due to their complex designs, these systems tend to be less mature and robust than widely used, time-tested file systems like Ext-4. Furthermore, the redesigned architecture for each media tier renders them incompatible with current file systems used by online systems, leading to high data migration costs for potential users.

The latest trend in utilizing NVM focuses on accelerating existing block device file systems. SPFS \cite{wooStackingPersistentMemory2023} and P2CACHE \cite{linP2CACHEExploringTiered2023} are two examples of NVM file systems layered on top of traditional disk file systems, aiming to enhance disk file systems by leveraging NVM. Generally, these approaches use NVM to speed up the slow persistence processes of disks. However, the speed improvements often come with certain side effects. For instance, SPFS introduces a second indexing overhead due to its overlay design and suffers from reduced re-access speed once data is absorbed by its NVM component. P2CACHE employs NVM to absorb not only synchronous writes but also asynchronous ones to ensure strong consistency, which unnecessarily slows down asynchronous writes. These drawbacks arise from their multi-tiered, stacked design, which fails to place NVM in its optimal position.

This poses a complex problem: \textbf{how to fully utilize the attractive characteristics of NVM while maintaining the advantages of the traditional software stack, ensuring compatibility and transparency for current user programs, and avoiding performance degradation and migration costs?} To address this, we propose NVLog, a transparent NVM write-ahead log designed to accelerate the performance of current disk file systems. Like SPFS and P2CACHE, NVLog aims to optimize the performance of existing mature disk file systems with NVM transparently. The unique contribution of NVLog lies in two aspects: (1) NVLog fully preserves the advantages of the DRAM page cache, ensuring that the use of NVM does not cause performance degradation in any use case; (2) NVLog is implemented as a log lies aside current VFS page cache, rather than an overlay file system, making it more efficient and compatible compared to previous work.

Though using faster media to accelerate slower ones has been a common practice in computer systems, things become different when it comes to NVM. 
The heterogeneity of DRAM, NVM, and disks \cite{wuRethinkingComputerArchitectures2016} presents significant challenges. First, these diverse devices differ in access granularity, speed, capacity, and price. Achieving optimal performance across all use cases requires leveraging the strengths and mitigating the weaknesses of each device, a task proven to be difficult by previous work \cite{linP2CACHEExploringTiered2023, wooStackingPersistentMemory2023}. Second, to fully exploit the performance of NVM, a meticulous design and implementation are necessary to prevent the software stack from becoming the primary bottleneck \cite{leeEmpiricalStudyNVMbased2018}. Finally, ensuring consistency across heterogeneous devices can be challenging, especially when writes are performed on different devices with varying timing and granularity.

NVLog records, and only records, synchronous writes into its NVM log structure, while preserving the DRAM cache to ensure that the performance of other normal operations remain unaffected. To maintain eventual consistency on the disk file system, the sequence of NVM syncs and disk write-backs is strictly defined. Additionally, NVLog optimizes the trigger mechanism of synchronous operations to avoid write amplification caused by scattered small writes. Furthermore, unlike previous work, NVLog's efficient write-ahead log design only requires a small portion of the NVM space, so the remaining free space can be utilized to support tiered caching or to serve applications that wish to use NVM directly.

The remainder of the paper is organized as follows. Section \ref{sec:bg} describes the characteristics of NVM and reviews related work on NVM. Section \ref{sec:motivation} presents our insights on heterogeneous storage system and the design principles of NVLog. Section \ref{sec:design} provides the detailed design of NVLog. Section \ref{sec:impl} discusses the implementation details. Section \ref{sec:evaluation} evaluates the performance of NVLog. Finally, Section \ref{sec:conclusion} concludes the paper. 

\section{Background}
\label{sec:bg}

\subsection{Non-volatile Memory}

Non-volatile memory (NVM) is a type of byte-addressable memory that can persist data. According to the JEDEC specification \cite{JEDEC}, NVM can be categorized into three types: NVDIMM-F, which uses flash storage on a DIMM; NVDIMM-N, which combines flash and DRAM on the same module, typically with a backup power source; and NVDIMM-P, which is inherently persistent as the computer's main memory, including technologies such as PCM, RRAM, and STT-RAM \cite{hosomiNovelNonvolatileMemory2005, IntelOptanePersistent, leeArchitectingPhaseChange2009, wongMetalOxideRRAM2012}. Intel Optane \cite{IntelOptanePersistent} has been the most popular NVM (NVDIMM-P) technology in recent years. Although Intel has discontinued Optane, alternative technologies have emerged in the market \cite{ComputeExpressLink2023, MSSSDSamsungMemory}, and we anticipate various substitutes in the future. In our work, NVM mainly refers to \textit{NVDIMM-P}, which is the most widely used technology, but our methods can also be applied to NVDIMM-N modules.

Typically, NVM offers intermediate performance, capacity, and price characteristics between DRAM and SSDs \cite{dulloorSystemSoftwarePersistent2014, izraelevitzBasicPerformanceMeasurements2019, xiangCharacterizingPerformanceIntel2022, wuRethinkingComputerArchitectures2016}, which presents both opportunities and challenges for interested software \cite{wuRethinkingComputerArchitectures2016, zhangDesignApplicationNew2023a}. Efforts to utilize NVM can be broadly categorized into three approaches.

The most efficient method is to expose NVM directly to user-space programs, allowing them to manipulate NVM with minimal software stack latency \cite{cuiSwapKVHotnessAware2023, volosAerieFlexibleFilesystem2014, wangPacmanEfficientCompaction2022, zhongMadFSPerFileVirtualization2023, maAsymNVMEfficientFramework2020, ruanPersistentMemoryDisaggregation2023, chenScalablePersistentMemory2021, dongPerformanceProtectionZoFS2019, jiFalconFastOLTP2023}. This approach is ideal for new programs designed to fully leverage NVM. \textit{However, for existing programs, particularly large and complex ones, migrating to new hardware with a new access pattern requires substantial effort, which can sometimes be impractical.}

A simpler concept is to use NVM as a slow second-tier memory for caching relatively cold data \cite{guoMiraProgramBehaviorGuidedFar2023, marufTPPTransparentPage2023, MigratePagesLieu, raybuckHeMemScalableTiered2021, renMTMRethinkingMemory2024, weinerTMOTransparentMemory2022}. This method takes advantage of NVM's price and capacity benefits, \textit{but does not leverage its persistence characteristics}.

A more balanced approach is to incorporate NVM file systems within the kernel. Since applications typically interact with persistent storage devices through kernel file systems, integrating NVM file systems can be seamlessly applied to existing applications.

\subsection{File Systems for NVM}

Various file systems have been proposed to exploit the performance of NVM devices. Previous work can be broadly categorized into two types: NVM-specialized file systems and cross-media file systems.

\noindent \textbf{NVM-specialized file systems:} Unlike traditional file systems designed for slower block storage devices like SSDs and HDDs, NVM-specialized file systems are tailored to leverage the unique characteristics of NVM. Direct Access (DAX) \cite{DAX} was an early attempt to bypass the DRAM page cache in traditional file systems. By using DAX, block file systems can operate directly on NVM without additional DRAM copies. DAX can be applied to various block file systems, such as Ext-4 and XFS. However, as a patch on block file systems, DAX does not fully exploit the byte-addressable nature of NVM. It still incurs unnecessary software overhead and lacks proper consistency guarantees. Despite these limitations, the concept of DAX has influenced subsequent work.

For some later efforts, NOVA \cite{xuNOVALogstructuredFile2016} is a dedicated file system for NVM, designed to leverage its fast and persistent characteristics. NOVA introduces a specially designed log structure on NVM to provide strong consistency and high performance. However, as shown in Figure \ref{fig:speed}, NOVA sometimes performs worse than traditional file systems with DRAM page cache due to the relatively lower speed of NVM. Additionally, its copy-on-write (CoW) design does not fully capitalize on NVM's byte-addressable access pattern. Besides, high-performance user-space (or user-kernel hybrid) file systems \cite{volosAerieFlexibleFilesystem2014, zhongMadFSPerFileVirtualization2023, chenScalablePersistentMemory2021, dongPerformanceProtectionZoFS2019, liuOptimizingFileSystems2024} have also been proposed to reduce latency caused by kernel traps but are not widely adopted due to compatibility issues with existing applications. \textit{These NVM-specialized file systems focus on persisting data on NVM but do not address the intermediate speed and capacity of NVM compared to DRAM and SSDs.
}

\noindent \textbf{Cross-media file systems:} Cross-media file systems introduce monolithic file systems with a holistic view over multiple heterogeneous storage tiers. These file systems are typically deployed across DRAM, NVM, SSD, and HDD. Strata \cite{kwonStrataCrossMedia2017} uses a user-space LibFS with per-process update logs to accelerate single-process access. Data is then digested into a shared area protected by the kernel and strategically placed into different media. Ziggurat \cite{zhengZigguratTieredFile2019} aims to accelerate slower disk storage with NVM, providing a file system with the speed of NVM and the capacity of larger disks. By predicting user write operations, Ziggurat dynamically directs operations to the appropriate storage tier. These cross-media file systems offer a comprehensive solution for balancing speed, capacity, and price by utilizing DRAM, NVM, and disk. \textit{However, the monolithic design of these systems makes them difficult to tailor and deploy.
Additionally, due to their complexity, a significant amount of time may be required for verification and modification before they reach maturity.}

\subsection{Enhancing Disk File Systems with NVM}
\label{sec:bg:diskacc}

A novel method to utilize NVM is to enhance existing disk file systems. The earliest attempts in this area involved transferring the journals of existing disk file systems to NVM. Since file systems with journaling incur 2.7 times higher write traffic than those without journals \cite{leeUnioningBufferCache2013}, some works \cite{JournalJbd2Linux, zhangCostEfficientNVMBasedJournaling2017} utilize NVM to absorb the commit and checkpoint operations of the journal, thereby reducing the write traffic to the disk. \textit{However, such approaches can only accelerate the journaling phase, and data still needs to be written to disk. Therefore, in synchronous scenarios, the overhead of direct disk I/O remains unavoidable.}

Another approach is to use NVM to transparently accelerate the synchronous writes of existing disk file systems, which are inherently slow due to their immediate persistence requirements and the slow speed of disk I/O. Although synchronous writes occur less frequently than normal read/write operations, they often become the main bottleneck of disk file systems, especially under workloads with strict correctness requirements, such as databases.

SPFS \cite{wooStackingPersistentMemory2023} is a stackable file system that lies on top of a traditional disk file system. Small synchronous write operations are directed to the overlaid NVM layer to eliminate I/O costs associated with slower disks, while other writes are handled by the lower file system. Since the page cache is maintained in the lower layer, non-synchronous read and write operations can still benefit from the speed of DRAM. SPFS takes into account the different performance characteristics of DRAM and NVM. However, SPFS relies on predictions to absorb synchronous writes, with these predictions based on the access patterns of previous synchronous writes. Before a successful prediction is made, the system still suffers from the low performance of synchronous I/O. This is an obvious problem when synchronous operations occur without a regular pattern. Additionally, once data is directed to the upper NVM layer, subsequent read operations also have to be performed on NVM, leading to unnecessarily high latency. Moreover, the two-layer design of SPFS introduces a double-indexing overhead, which lowers the overall performance of the system. 

P2CACHE \cite{linP2CACHEExploringTiered2023} is another overlay file system. There are two key differences between P2CACHE and SPFS. First, P2CACHE positions the DRAM cache to the same level as the NVM layer. Any data is written to both NVM and DRAM simultaneously, allowing P2CACHE to serve subsequent read operations from the faster DRAM. Second, P2CACHE absorbs not only synchronous but also normal write operations to its NVM layer to provide strong consistency. Although P2CACHE claims to benefit from overlapping writes between NVM and DRAM, for normal writes, writing to NVM is still slower than writing to DRAM. 
Furthermore, as an experimental work, P2CACHE fails to support several necessary functions, such as page status management, page locking, and reclamation support, which are indispensable for a real operating system. Additionally, implementing page cache as non-volatile also requires changes from applications, as they previously assumed that the page cache was volatile.

\textit{Overall, these overlay file systems fail to put NVM into the proper position of the memory hierarchy and do not fully exploit its advantages, as we will illustrate in Section \ref{sec:motivation}, resulting in limited acceleration.
}

\section{Motivation}
\label{sec:motivation}

\begin{figure}
    \centering
    \includegraphics[width=\linewidth]{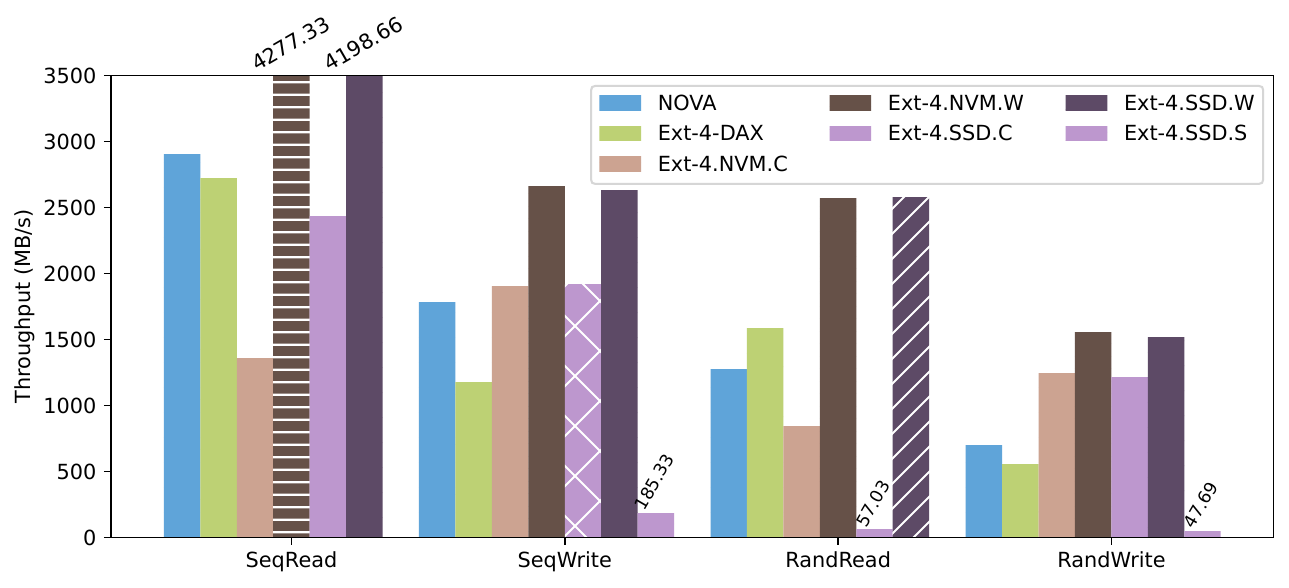}
    \caption{The throughput on different file systems and different storage devices, tested with FIO. C and W suffixes indicate that the page cache is cold (cache miss) or warm (cache hit). S means sync writes. Reads are not affected by sync. }
    \label{fig:speed}
\end{figure}

In current operating systems, storage devices are typically managed by file systems. We evaluated the basic performance of a traditional disk file system, Ext-4, with and without page cache, and compared it with file systems on NVM (NOVA, Ext-4, and Ext-4-DAX).
The testbed of this evaluation is the same as that of Section \ref{sec:evaluation}.
Figure \ref{fig:speed} shows that operations performed on the DRAM page cache can always achieve higher performance than those on NVM. \textit{The main limitation of current disk file systems lies in sync writes and cache-missing operations.} We also performed a breakdown of the access latency. The results show that the slowdown of these cache-absent operations is primarily due to synchronous I/O and additional software stack overheads. For cache-missing reads and sync writes, the direct I/O requirement is the main contributor (over 90\%) to the slowdown. For cache-missing writes, although data can be written to the DRAM cache without I/O, memory allocation and index building for the absent pages become the primary factors (70\%) of performance degradation. 

The observation is simple and intuitive: performance degradation occurs only when a faster storage layer, in this case, the DRAM cache, is missing. Therefore, we believe that preserving an efficient, high-speed DRAM cache, while accelerating cache-missing paths—sync writes and cache-absent operations—with NVM, presents a promising approach to further enhancing the current storage software stack.

Sync writes are common in strong-consistency workloads, such as databases, which form the backbone of many Internet applications. In such scenarios, we aim to provide acceleration similar to that of NVM file systems—by absorbing foreground sync writes to faster NVM instead of writing to disk—thereby optimizing throughput and latency. However, in cases where sync writes are infrequent, NVM file systems are less competitive than disk-based file systems because they fail to efficiently utilize DRAM. In these situations, we strive to maintain the fast DRAM path while also accelerating slow sync writes, thus providing further performance improvements rather than degradation.

For cache-missing async operations, extending the size of the page cache can be beneficial. Methods \cite{guoMiraProgramBehaviorGuidedFar2023, marufTPPTransparentPage2023, MigratePagesLieu, raybuckHeMemScalableTiered2021, renMTMRethinkingMemory2024, weinerTMOTransparentMemory2022} have been proposed to extend the DRAM cache with large and cheap NVM, known as tiered memory; however, they usually do not integrate well with NVM file systems, as these file systems often require a large NVM space. We aim to further support tiered memory by reducing our NVM usage.

\textbf{In summary, the goal of this work is to use NVM to accelerate the current storage system efficiently and transparently, without any performance downgrade or data/code migration, while requiring only minimal NVM persistent space, allowing the remaining space to be used for other purposes such as extending the page cache.} 
We hence introduce NVLog, an NVM-based write-ahead log integrated within the VFS page cache, to efficiently accelerate the slow sync writes of current disk file systems while maintaining transparency to user applications and compatibility with the system storage software stack. 

NVLog is based on two insights that were ignored by previous disk file system accelerators (Section \ref{sec:bg:diskacc}): 

\begin{description}
\item[I1:] The DRAM cache is sufficient and efficient to serve applications.
Therefore, when persisting synchronized data to NVM, the focus should be on the efficiency of \textit{recording}, rather than \textit{data retrieval}. 

\item[I2:] Establishing a well-defined write timing between NVM and disk is crucial for ensuring crash consistency while minimizing the amount of data written to NVM. 
\end{description}

Due to neglecting \textbf{I1}, both P2CACHE \cite{linP2CACHEExploringTiered2023} and SPFS \cite{wooStackingPersistentMemory2023} have to create an index for data on NVM for subsequent reads, and have difficulty reducing the space usage on NVM. Due to neglecting \textbf{I2}, these systems are forced to also redirect async writes to NVM when absorbing sync writes, in order to avoid inconsistencies between the data from sync writes (to NVM) and async writes (to disk).

To respond to the insights and achieve our goals, NVLog is designed and implemented with four principles:

\begin{description}

\item[P1:] \textbf{Transparency to both upper applications and lower file systems.} NVLog should function as a transparent file system accelerator, requiring no changes to applications to benefit from the acceleration. Additionally, NVLog should not necessitate any modifications to the underlying file systems, allowing all time-tested disk file systems to be accelerated without additional cost. This downward transparency also means that expensive data migration is not required for users.

\item[P2:] \textbf{No consistency change to the current I/O stack.} NVLog should not alter the current consistency model. Increasing the consistency level would incur higher costs without necessarily benefiting existing applications, while decreasing it could compromise the semantics of current file systems. NVLog adheres to the belief that the existing consistency level represents the best practice for current applications.

\item[P3:] \textbf{No performance downgrade to current file systems.} The goal of NVLog is to address the limitations of disk file systems rather than provide unbalanced performance like previous works \cite{dulloorSystemSoftwarePersistent2014, xuNOVALogstructuredFile2016, linP2CACHEExploringTiered2023, wooStackingPersistentMemory2023}. Therefore, NVLog should ensure that it does not degrade the performance of underlying file systems under any circumstances.

\item[P4:] \textbf{Lightweight design, minimal persistence footprint. } To ensure stability, the design of NVLog is kept as simple as possible. Additionally, NVLog occupies only the minimum necessary NVM space, allowing the remaining portion to be used for tiered caching or other optimization methods to further enhance system performance.

\end{description}

\section{NVLog Design}
\label{sec:design}

\begin{figure}
    \centering
    \includegraphics[width=0.8\linewidth]{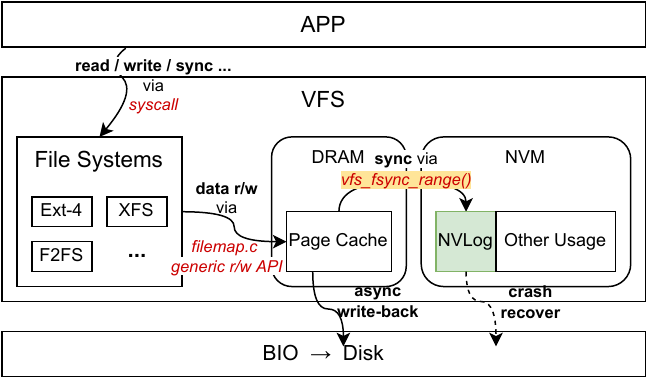}
    \caption{NVLog Architecture. Figure shows the position and data flow of NVLog inside the Linux kernel.}
    \label{fig:arch}
\end{figure}

NVLog is designed as a write-ahead log that exclusively absorbs sync writes, while all other normal read/write operations are still served by the DRAM page cache. Figure \ref{fig:arch} illustrates the position and data flow of NVLog inside the Linux kernel. In this section, we first present the log structure design and the write process of NVLog. Second, we introduce sync semantic optimization to reduce write amplification in small, scattered sync writes. Then, we discuss the crash consistency challenges and solutions faced by NVLog. Finally, we cover the crash recovery and garbage collection processes.

\subsection{Log Structure}

\begin{figure}[t]
    \centering
    \includegraphics[width=0.9\linewidth]{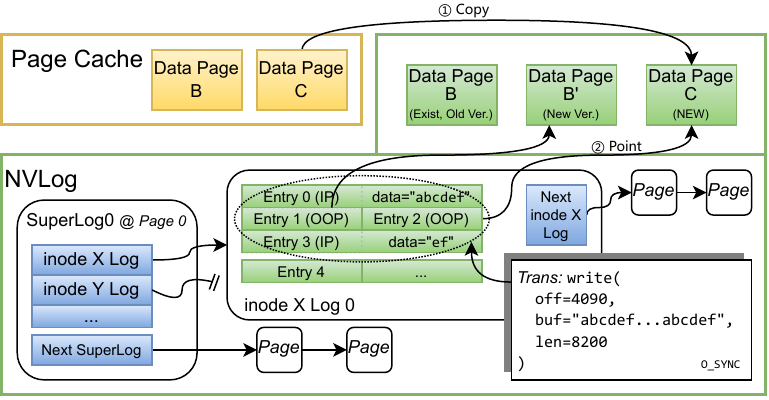}
    \caption{Log Structure. The detailed design of the data arrangement and the behavior of sync writes. The yellow parts are data pages in DRAM cache, the green parts are persistent data logs in NVM, and the blue parts are persistent metadata that helps NVLog to persist data. }
    \label{fig:log}
\end{figure}

As depicted in Figure \ref{fig:log}, the fundamental design of NVLog consists of a series of logs on NVM. This log-based design offers two major benefits to NVLog: First, it eliminates the need for indexing, allowing for fast append-only writes (recall \textbf{I1}), which ensures high performance. Second, out-of-date data can be easily recycled, enabling NVLog to temporarily use only a small portion of the NVM space (recall \textbf{P4}). Both of these advantages have not been adopted in previous work \cite{linP2CACHEExploringTiered2023, wooStackingPersistentMemory2023}, making NVLog a novel approach in leveraging NVM for file system acceleration.

\subsubsection{Memory Arrangement}

For each log series (inode log/super log, introduced in Section \ref{sec:design:logtype}), the 64B log entries are first sequentially organized in a 4KB page. When the log exceeds the current page, another page is allocated. This new page is then linked to the previous page by a linked list, creating a chain of log pages. The traversal operation on the log is accelerated by prefetching.

\subsubsection{Log Type}
\label{sec:design:logtype}
There are two types of logs in NVLog. The first type is the \textit{super log}, which contains pointers that link to the log heads of all the inodes managed by NVLog. 
The second type of log is the \textit{inode log}. Each file managed by NVLog has its own inode log, where all sync writes and metadata updates are recorded. 
For example, in Figure \ref{fig:log}, the first entry in the super log, marked as \texttt{inode X Log}, maintains a pointer to the head log page of inode X, which is \texttt{inode X Log 0}. Subsequent log pages for inode X are linked together via a linked list. Each inode X log page contains plenty of synchronous update events (green squares in the figure) related to inode X.

There is only one global super log in NVLog, and its first log page (log head) is located at the physical 0 address of the NVM device. This placement ensures that NVLog can easily locate the super log directly after a power failure. The super log serves as the root of all logs, from which all persistent domain data can be found and replayed.

\begin{figure*} [ht]
    \centering
    \includegraphics[width=\linewidth]{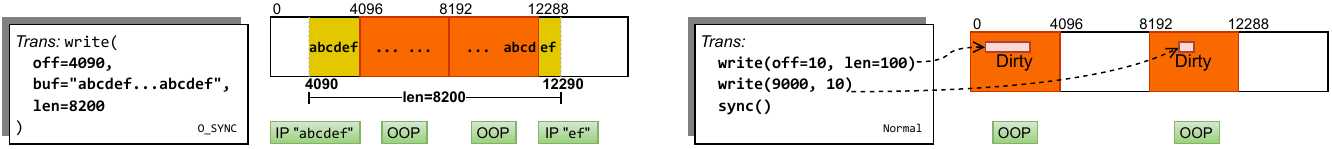}
    \caption{Sync Write Step. For \texttt{O\_SYNC} writes, each write operation is divided into parts by page boundaries. Page-aligned (red) parts are recorded by OOP entries, and the unaligned (yellow) parts are recorded by IP entries. For normal writes followed by an \texttt{fsync}, all dirty pages are recorded by OOP entries. }
    \label{fig:trans}
\end{figure*}

\subsubsection{Log Entry}
\label{section:entry}

Each entry in the super log describes an inode log. The structure of the super log entry looks as follows: 
\begin{verbatim}
struct superlog_entry
{
    dev_t           s_dev;
    unsigned long   i_ino;
    uint32_t        head_log_page;
    struct inodelog_entry *committed_log_tail;
}
\end{verbatim}

The \texttt{s\_dev} and \texttt{i\_ino} fields are used to locate a specific file; the \texttt{head\_log\_page} pointer points to the first log page of this inode log; the \texttt{committed\_log\_tail} field indicates the current log tail of the inode. Whenever a new inode is delegated to NVLog, a new super log entry is created, pointing to the new inode log. Then when this file requires recovery, NVLog retrieves its inode log by finding the corresponding super log entry. Note that the DRAM inode structure also holds a pointer to its NVLog log head, eliminating the need to search for its super log entry, so there's no extra overhead during regular access.

The inode log captures all synchronous operations performed on the inode. Since NVLog manages NVM in pages, writes are naturally divided into segments no larger than a page. Consequently, if a write operation spans across $n-1$ page boundaries, it may need to be recorded up to $n$ times. Each entry in the inode log corresponds to a write segment, and the format of each entry is as follows:
\begin{verbatim}
struct inodelog_entry
{
    uint16_t    flag;
    uint64_t    file_offset;    
    uint16_t    data_len;
    uint32_t    page_index;
    uint64_t    last_write;
    uint64_t    tid;
}
\end{verbatim}

There are two types of inode log entries: out-of-place log entries (OOP entries) and in-place log entries (IP entries). In an OOP entry, the data is stored in a separate page pointed by the \texttt{page\_index} field. In contrast, an IP entry stores the data within the log zone itself. The type of entry is distinguished by the \texttt{page\_index} field: if \texttt{page\_index} is 0, the entry is an IP entry; otherwise, it is an OOP entry.

NVLog uses the two types of entries to record data of different lengths. For whole-page data, NVLog employs OOP entries to perform large, page-aligned shadow-paging writes. On the one hand, this approach facilitates data page allocation and reclamation. On the other hand, as the new OOP data page is filled with fresh data, there's no need to copy the old data in write process (e.g., \texttt{Entry 1} and \texttt{Data Page B'} in Figure \ref{fig:log}). For smaller, unaligned write segments, which can have arbitrary sizes, NVLog uses IP entries to record them according to their actual length (e.g., \texttt{Entry 0} and \texttt{Entry 3} in Figure \ref{fig:log}). By leveraging the byte-addressable nature of NVM, this method helps to avoid write amplification.

Except for the differences mentioned above, OOP entries and IP entries manipulate the remaining fields in the same way. The \texttt{file\_offset} field indicates the position in the file that the current entry is writing to. The \texttt{data\_len} field specifies the length of the current write. The \texttt{last\_write} field is reserved for locating the previous write at the same position during backtracking (see Section \ref{sec:recover}). The \texttt{tid} field marks a transaction. The \texttt{flag} field indicates the state of an entry.

Apart from the two types of write entries, there is also a metadata update entry to record changes to the inode's metadata, and a write-back record entry, which will be discussed in Section \ref{section:hconsist}. Their structures are straightforward and can be found in our open-source code, so we will not list them here.

\subsection{System Modification}
\label{section:steps0}

Let's first examine the sync write steps of the Linux kernel. When a write is performed, the application first uses a \texttt{syscall} to invoke the file system. The file system then writes the data to the DRAM page cache and marks the relevant pages as dirty. Later, when a sync is performed, \texttt{vfs\_fsync\_range} is called to synchronously write the dirty pages back to the disk. If no sync is performed, the dirty pages will eventually be written back to the disk asynchronously in the background.

Previous works intercept the \texttt{syscall} by overlaying an NVM file system, thereby achieving write absorption on lower disk file systems. However, this approach separates the NVM accelerator from the system page cache, leading to double indexing overhead and inefficient redundant operation redirection. In contrast, NVLog absorbs sync writes within \texttt{vfs\_fsync\_range}. This allows NVLog to focus solely on absorbing sync writes while maintaining the benefits of the DRAM page cache provided by the system, resulting in higher performance (recall \textbf{P3}) and lower resource usage (recall \textbf{P4}).

It is important to note that NVLog does not alter the existing functions of the DRAM page cache. Writing data to NVLog does not clear the dirty flag of the page. NVLog simply converts synchronous disk write-backs into asynchronous ones, and uses NVM to perform immediate data persistence. 
This ensures that all data will eventually appear on the disk. 
Meanwhile, an extra flag is added to track the dirty pages that have been absorbed by NVLog. This ensures that the same write will not enter NVLog multiple times.

By converting sync writes into periodical async writes, NVLog also provides some additional benefits. For \texttt{fsync} operations, each write to disk typically involves a data write and a metadata write. NVLog can first convert both of these writes into NVM writes and later, during an async write-back, aggregate the multiple metadata updates caused by the data updates, writing them to disk in one operation. This reduces the write pressure on the disk and extends its lifespan (especially for SSDs). For \texttt{fdatasync} operations, if it's an append write, new disk blocks also need to be allocated through metadata operations. Similarly, NVLog can optimize block allocation by aggregating multiple synchronous writes over a period of time to improve contiguous block allocation.

\subsection{Sync Write Steps}
\label{section:steps}

Sync writes can be categorized into two types: \texttt{O\_SYNC} writes and \texttt{fsync}-like calls (e.g., \texttt{fsync} and \texttt{fdatasync}). These two kinds of sync leads to different system behaviors, as shown in Figure \ref{fig:trans}. For normal writes, the behavior is straightforward for NVLog: record all dirty pages to NVM. To better illustrate the NVLog write process, we will take the more complex \texttt{O\_SYNC} writes as an example.

To respect \textbf{P2}, each sync write operation is regarded as a transaction in NVLog, and is assigned an auto-increment id. The write transaction is broken into segments according to the page boundaries it crosses. For each segment, a log entry is appended to the end of the inode log of the current file. Aligned whole-page segments are recorded by OOP entries, with the data copied from the DRAM cache to a newly allocated NVM data page (e.g., \textcircled{1} in Figure \ref{fig:log}). Unaligned segments are recorded by IP entries, with the data copied to the entry's data zone. Entry bodies are filled according to the definition in Section \ref{section:entry}. Note that even if there is a previous OOP NVM data page found for the same offset (e.g., \texttt{Data Page B} in Figure \ref{fig:log}), we cannot reuse it. Otherwise, it might result in the loss of data from the previous transaction if a crash occurs before the current transaction ends. 

To further ensure \textbf{P2}, several techniques are applied. First, the \texttt{committed\_log\_tail} of the inode log is only updated atomically after all the segments in a transaction have finished, ensuring its integrity. 
Second, due to the presence of CPU caches, writes to NVM may return before they are eventually persisted to the NVM device. To eliminate this inconsistency risk, NVLog uses cache line write-back (e.g., Intel's \texttt{clwb}) to explicitly instruct the CPU to flush data back to NVM. If the system support eADR \cite{EADRNewOpportunities}, the cache line write-back process can be omitted, allowing NVLog to achieve better performance. 
Third, memory barriers (\texttt{sfence}) are employed to maintain the ordering of store operations before and after consistency-critical points. In NVLog, only two barriers are used. The first is placed after all transaction segments are logged and before the \texttt{committed\_log\_tail} is updated, ensuring that the transaction is complete before it can be seen. The second barrier is placed after the commit and before the start of the next transaction to maintain order between transactions.

\algdef{SE}[VARIABLES]{Variables}{EndVariables}
   {\algorithmicvariables}
   {\algorithmicend\ \algorithmicvariables}
\algnewcommand{\algorithmicvariables}{\textbf{global variables}}
\algdef{SE}[]{MultiProcedure}{EndMultiProcedure}
   {\algorithmicmultproc}
   {\algorithmicend\ \algorithmicmultproc}
\algnewcommand{\algorithmicmultproc}[1]{\textbf{procedure} #1}
\algnewcommand{\LineComment}[1]{\State \(\triangleright\) #1}

\begin{algorithm}
\caption{Active sync mechanism. }
\label{alg:act_sync}

\begin{algorithmic}[1]

\Variables
    \State $should\_active\_cnt$ \Comment{Counter before activate. }
    \State $should\_deact\_cnt$ \Comment{Counter before deactivate. }
\EndVariables

\MultiProcedure{\textsc{MarkSync}(\parbox[t]{\linewidth}{$file$, $written\_bytes$, \\$dirty\_pages$, $sensitivity$)}}
    \LineComment{This is called on each sync. }

    \If{$written\_bytes < dirty\_pages * 4096$}
        \State $should\_active\_cnt \gets should\_active\_cnt + 1$
        \If{$should\_active\_cnt \ge sensitivity$}
            \State $file.flags \gets file.flags \,|\, O\_SYNC$
            \State $should\_deact\_cnt \gets 0$
        \EndIf
    \EndIf

\EndMultiProcedure

\MultiProcedure{\textsc{ClearSync}(\parbox[t]{\linewidth}{$file$, $written\_bytes$, \\$dirty\_pages$, $sensitivity$)}}
    \LineComment{This is called on each write. }

    \If{$written\_bytes \ge dirty\_pages * 4096$}
        \State $should\_deact\_cnt \gets should\_deact\_cnt + 1$
        \If{$should\_deact\_cnt \ge sensitivity$}
            \State $file.flags \gets file.flags \,\&\, {\sim} O\_SYNC$
            \State $should\_active\_cnt \gets 0$
        \EndIf
    \EndIf

\EndMultiProcedure

\end{algorithmic}
\end{algorithm}

\subsection{Active Sync Optimization}
\label{section:actsyn}

As depicted in Figure \ref{fig:trans}, \texttt{O\_SYNC} and \texttt{fsync} exhibit different behaviors. \texttt{O\_SYNC} is a flag indicating that a file or a mount point should always be written synchronously. Therefore, sync can be performed within the write syscall, at which point the exact range of data that needs to be recorded in NVLog is known. In contrast, the \texttt{fsync} operation is a post-write instruction, which means that at this point, we only know the dirtied data in the granularity of page instead of bytes. Consequently, if small, scattered writes occur before an \texttt{fsync}, the sync operation will write all the dirtied full pages to the NVM, resulting in severe write amplification. For example, in Figure \ref{fig:trans}, the pink fragments will cause all contents of the red dirty pages to be written to NVM.

To address the write amplification problem of \texttt{fsync}, we introduce \textit{active sync}. The main idea is to predict whether subsequent synchronous writes will be more efficient if performed on full pages or at the byte level. Since application access patterns typically exhibit temporal locality, we can use past access patterns to predict future behavior. We track the count of dirtied pages and the number of written bytes between two sync operations, and then compare them according to Algorithm \ref{alg:act_sync}. Based on this comparison, we can dynamically apply or withdraw the \texttt{O\_SYNC} flag on files that are not originally marked as \texttt{O\_SYNC}.

Taking Figure \ref{fig:trans} \texttt{Normal} as an example, the total written byte count is 110, and the dirtied page count is 2. In \texttt{Normal} mode, the sync operation results in 4096*2 bytes being written to NVLog. However, if it were in \texttt{O\_SYNC} mode, only 110 bytes would need to be written. Based on this observation, we can predict that the performance of this file may be better in \texttt{O\_SYNC} mode in the near future, so we proactively mark it as \texttt{O\_SYNC}. Note that \texttt{written\_bytes} may larger than \texttt{dirty\_pages*4096} because there can be repeated writes to a single page.

The \texttt{sensitivity} parameter is used to tune the algorithm for different workloads to prevent thrashing. Our tests show that, unless the sync access pattern of the application is highly irregular, setting this value to 2 generally enhances the performance of small sync writes for most daily applications.

\subsection{Consistency between NVM and Disk}
\label{section:hconsist}

We now have two separate write paths to NVM and disk. Synchronous writes are persisted directly to NVLog and may involve fewer bytes than a full page. In contrast, any write operation, whether synchronized or not, triggers asynchronous whole-page write-backs from DRAM to disk. 

Pages on the underlying disk of NVLog always contain the correct data with the appropriate intra-page write sequences, because a disk page is written as a checkpoint of the page in DRAM cache, and the file system always ensures that the cached page is accurate.
However, since we have converted all sync writes to asynchronous writes on disk, there is a risk that the disk page might persist an older version of the data if a power failure occurs before a synchronized dirty page is written back to the disk. Fortunately, NVLog ensures that a fresher version of the data is persisted on NVM. Nevertheless, because we only persist the necessary bytes for sync writes and not all bytes of all dirty pages, there is a potential for inconsistency issues, which could violate \textbf{P2}.

\begin{figure}[t]
    \centering
    \includegraphics[width=\linewidth]{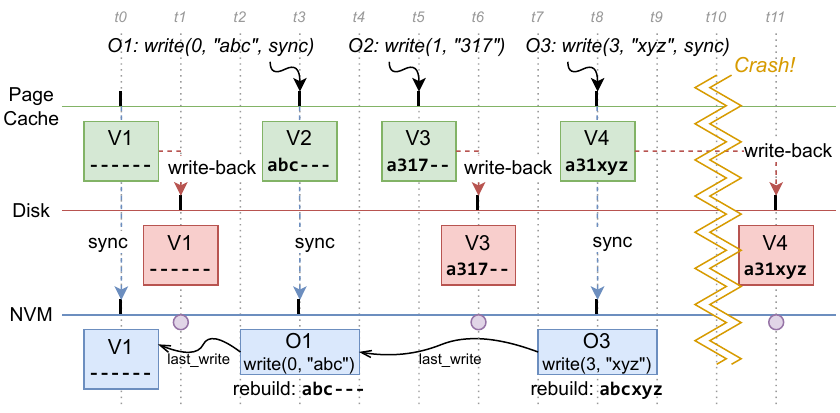}
    \caption{Page consistency on heterogeneous devices.}
    \label{fig:consistency}
\end{figure}

Figure \ref{fig:consistency} illustrates a typical scenario of inconsistency between the data on disk and NVM. At time $t2$, the page cache, disk, and NVM are all in a consistent state with version $V1$. By time $t7$, the disk holds the latest version $V3$, while NVM contains the previous version $V2$, which can be reconstructed from $V1$ and recorded $O1$. The discrepancy arises because $O2$ is not a sync write, and therefore, it is not persisted to NVM. The first issue is that if a crash occurs at $t7$, the recovery process will reconstruct $V2$ from NVM and overwrite the $V3$ on the disk, leading to an unexpected data rollback. 

A more critical issue arises at $t10$. At this point, a new sync write $O3$ has been performed and successfully recorded by NVM, but the new data $V4$ has not yet been written back to the disk. In the event of a crash here, we would only be able to rebuild \texttt{abcxyz} from NVM, which does not reflect any expected version. The data has been messed up on NVM at this point, yet the disk only has the previous version $V3$ instead of the latest sync version $V4$. 

The key to these frustrating problems is that we don't have the exact time sequence of disk writes and NVM writes. To address these problems, we persist disk write-back events on NVM (indicated by the purple bubbles on the NVM timeline in Figure \ref{fig:consistency}) to maintain a global clock. Whenever a disk write-back occurs, if (and only if, for the sake of performance) a valid previous entry exists, a write-back entry is appended to the inode log, marking the previous writes on this page as expired. During recovery, only unexpired operations are replayed to the page version on the disk. For example, at $t10$, we have $O1$ and $O3$ on the NVM, and $V3$ on the disk. The write-back entry (purple bubble) between $O1$ and $O3$ indicates that $O1$ and its predecessors are expired. So we only replay $O3$ to $V3$, resulting in \texttt{a31xyz}, which is the correct version reflecting the lost $V4$. This mechanism ensures that NVLog will always rebuild the latest data without rollback or errors.

It is worth mentioning that this algorithm is a key factor in enabling NVLog's performance to outperform previous works (recall \textbf{I2} and \textbf{P3}). Without this algorithm, absorbing only sync writes to NVM while maintaining the DRAM-disk path for other normal writes would result in confusion between NVM and disk data versions. To avoid this confusion, for example, P2CACHE absorbs all writes to NVM, thus leading to unnecessary performance degradation under asynchronous writes.

\subsection{Crash Recovery}
\label{sec:recover}

To recover the data on NVLog to the disk after a power failure, a crash replay procedure is introduced. This procedure involves multiple passes through each inode log to recover each file. First, it traverses the entire inode log. During this pass, log entries associated with the same data page offset are linked together in sequence via the \texttt{last\_write} field of each entry, and the latest write entry for each data page is saved into an index. After the whole index is built, the replay starts. For each page, the rebuilder walks from its latest entry back to the earliest via the \texttt{last\_write} field. Once a write-back entry or an OOP entry is found, the walk halts because the previous data are either expired or overwritten. The data from the traversed entries are then replayed to the data page on the disk.

Note that our scanning ends at \texttt{committed\_log\_tail}, and any uncommitted partial writes after this point will be dropped. This ensures that even though a single write might span multiple pages, it will be recovered in an all-or-nothing manner.
Also note that NVLog does not conflict with the journaling mechanism (e.g., JBD) of the file system, as NVLog only records events, not data blocks. After a crash, as long as the underlying file system remains intact, NVLog enables the file system to catch up to the expected status. Therefore, in such cases, running \texttt{fsck} should be the first step, followed by NVLog recovery. 
Since the recovery process only occurs after a crash reboot, it does not affect the performance of regular operations. We conducted various crash experiments on the experimental platform described in Section \ref{sec:evaluation}, and the results show that the recovery time is usually around 10 seconds.

As we can see, NVLog postpones the index-building process to the recovery period, whereas previous works maintain their index during regular runtime. The elimination of runtime indexing is another factor contributing to our lightweight and efficient design (recall \textbf{I1}).

\subsection{Garbage Collection}

To reclaim NVM space from NVLog and respond to \textbf{P4}, a garbage collector is provided as a kernel thread. This garbage collector periodically runs in the background, walking through the log pages to check for any log or data pages that are no longer needed. A log entry becomes obsolete when it expires due to a subsequent write-back or is overwritten by a later OOP entry. A data page is considered useless if its associated log entry is obsolete. A useless data page is reclaimed as soon as it is identified, and a log page is reclaimed once all its entries are deemed useless. The walk stops before the latest log page of each inode, as the latest page is obviously still in use. This scanning process does not require any locks, ensuring that it does not interfere with foreground operations.

Note that emerging NVM technologies, such as RRAM and MRAM, may have smaller capacities compared to PCM, which could lead to NVLog encountering situations where the NVM is full. In such cases, NVLog will fall back to performing sync writes to disk and wait for the GC to free up available NVM pages. NVLog does not need to wait for GC to complete; as long as there are free pages, it can resume using the NVM sync path.

\section{Implementation}
\label{sec:impl}

NVLog is implemented in the Linux kernel 5.15 (LTS) with 7.3K lines of kernel code and no more than 1K lines of code for auxiliary tools. The changes to the kernel code are limited to the Virtual File System (VFS) (0.3K LOC), the memory management system (6.2K LOC), and the drivers (0.8K LOC). 

The VFS code is minimally modified to transfer necessary events to NVLog without altering the current API semantics, thus maintaining full compatibility and transparency for existing applications and file systems (recall \textbf{P1}). The memory management system (MM) contains the majority of our work. All functions of NVLog discussed in Section \ref{sec:design} are implemented in MM. Meanwhile, extra page flags are added to track page status, and the mark/clear-dirty and write-back functions are modified to handle the new flags. A new driver module is added to initialize DAX NVM devices and configure NVLog. Furthermore, some utilities are provided to initialize, configure, and monitor NVLog from user space. The prototype is open source and can be found at \url{https://github.com/BugJLU/NVLog}.

\section{Evaluation}
\label{sec:evaluation}

\begin{figure*}[t]
    \centering
    \subfloat{\includegraphics[height=1.16in]{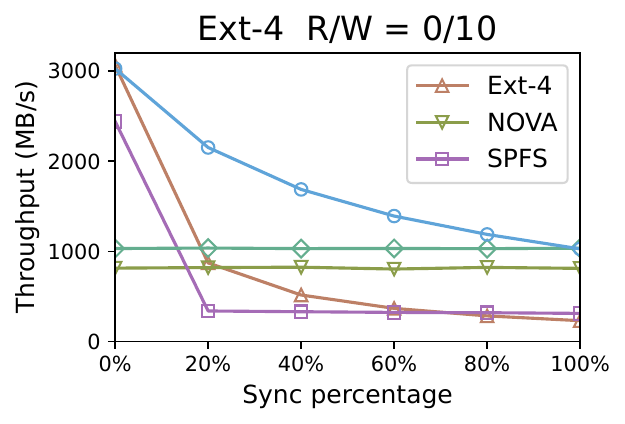} \label{fig:e1_e1}}
    \subfloat{\includegraphics[height=1.16in]{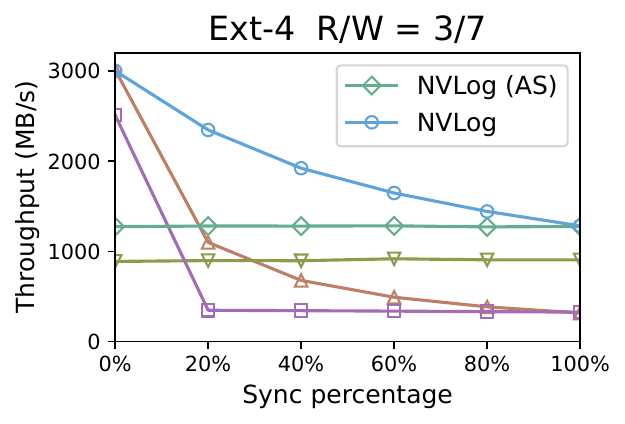} \label{fig:e1_e2}}
    \subfloat{\includegraphics[height=1.16in]{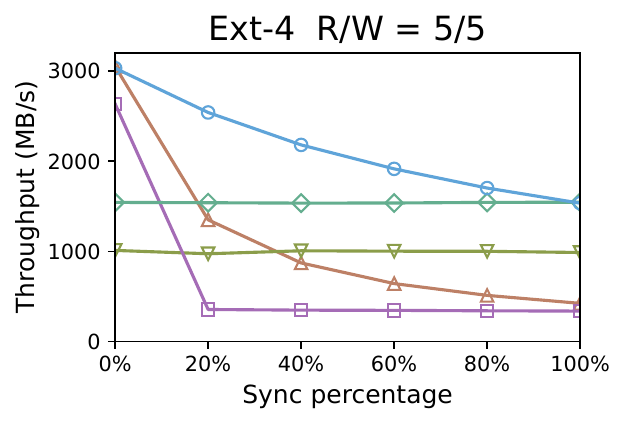} \label{fig:e1_e3}}
    \subfloat{\includegraphics[height=1.16in]{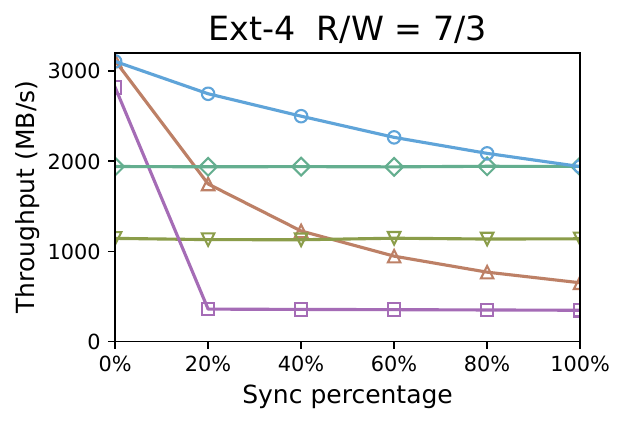} \label{fig:e1_e4}}
    \\
    \subfloat{\includegraphics[height=1.16in]{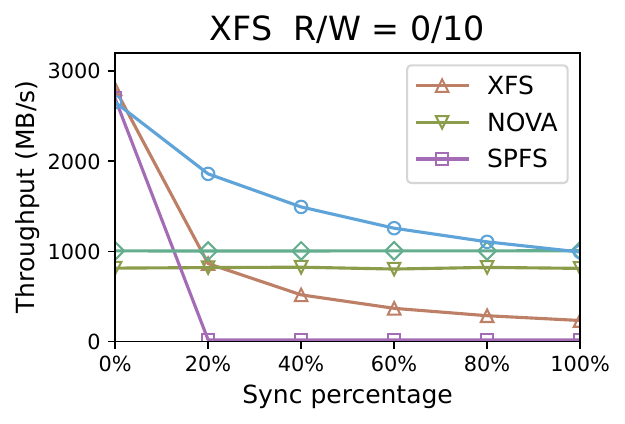} \label{fig:e1_x1}}
    \subfloat{\includegraphics[height=1.16in]{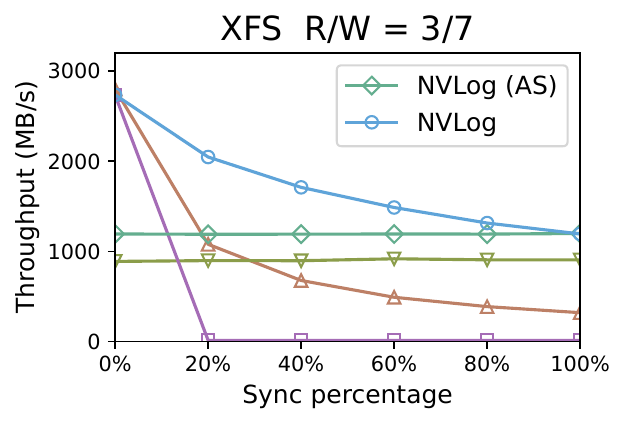} \label{fig:e1_x2}}
    \subfloat{\includegraphics[height=1.16in]{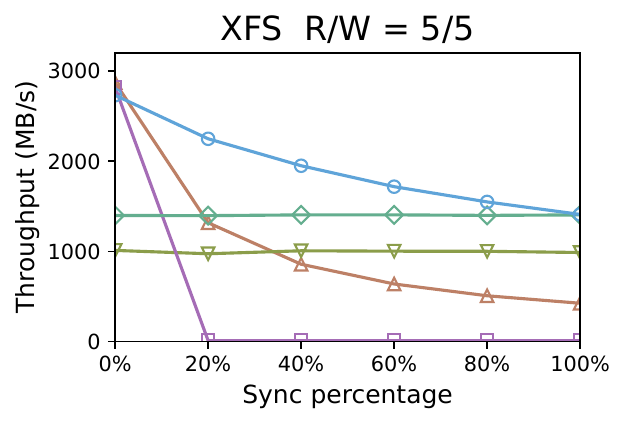} \label{fig:e1_x3}}
    \subfloat{\includegraphics[height=1.16in]{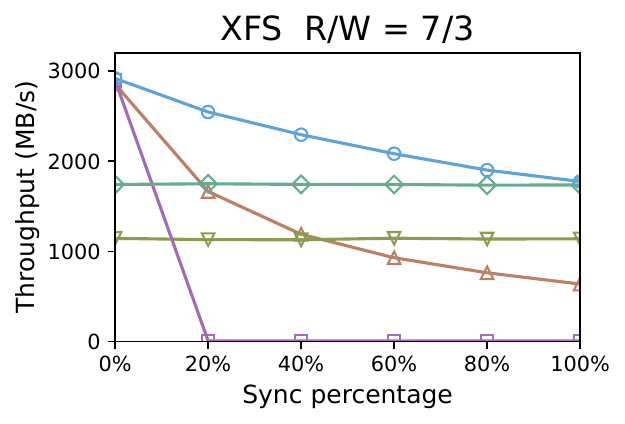} \label{fig:e1_x4}}
    \caption{Read, write, and sync mixed tests under 4KB random access. AS: all writes are forced to be synchronized.}
    \label{fig:exp:ratio}
\end{figure*}

In this section, we evaluate NVLog using micro- and macro-benchmarks, to demonstrate the performance advantages and limitations against other file systems. We compare NVLog with NOVA and SPFS, which respectively represent the state-of-the-art work in NVM file systems and NVM overlay accelerators. We did not select P2CACHE as the representative because its open-source code is incomplete to run properly. Unless specified, file systems and workloads use default configurations (e.g., ordered-journaling for Ext4).

All the experiments are conducted on a system equipped with Intel Xeon 5218R (20 cores), 128GB DRAM, 256GB Intel Optane PMEM (128GB x2, interleaved), Samsung PM9A3 1.92TB NVMe SSD, and Ubuntu 20.04. Note that the SSD used in the evaluation has a high speed, whereas the bandwidth of the NVM is limited because only two modules are installed. Therefore, our experiments roughly represent the lower bound of acceleration that NVLog can bring. In systems with slower storage (e.g., SATA SSDs or HDDs) and higher bandwidth NVM (e.g., more PMEM modules installed), the performance improvement ratio of NVLog will be much higher than the results reported in this section.

\subsection{Microbenchmarks}

We use multiple microbenchmarks to evaluate the performance of NVLog and other file systems. In this section, we choose both Ext-4 and XFS as baselines to demonstrate the flexible adaptability of NVLog. Similar to previous work, NVLog does not accelerate cold I/O from disk, as most workloads benefit from caching. Therefore, to better demonstrate the advantages of NVLog's design in common use cases, our experiments are conducted with data pre-loaded into the cache (by reading the test file for one pass). 

\subsubsection{Mixed Operation Access}

To comprehensively illustrate the performance advantages of NVLog's design, we first deploy 4KB random read/write tests with varying r/w ratios (0/10, 3/7, 5/5, 7/3) and different sync write percentages (from 0\% to 100\% in steps of 20\%). NVLog can absorb sync operations on demand due to its consistency design described in Section \ref{section:hconsist}. To highlight the impact of this design, we also test NVLog (AS, always-sync) in this experiment, representing performance without the aid of the consistency design, where all writes must be persisted to NVM to ensure consistency, similar to the strategy used by P2CACHE.

The experimental results are shown in Figure \ref{fig:exp:ratio}. Thanks to our DRAM-NVM cooperative design, NVLog outperforms NVM FS, disk FS, and NVM-based FS accelerators in most cases. In non-sync workloads, by leveraging the DRAM page cache, NVLog performs similarly to its baseline disk FS, achieving speeds up to 3.72x, 2.93x, and 1.24x faster than NOVA, NVLog (AS), and SPFS, respectively. In partial-sync workloads, NVLog outperforms the disk FS, NOVA, and SPFS by up to 4.44x, 2.62x, and 324.11x, respectively. In this test, SPFS demonstrates poor performance.
Our performance breakdown shows that 97\% of its time is spent on indexing, indicating that its indexing mechanism suffers from significant degradation in random access scenarios. The results show that NVLog consistently maintains a good balance between DRAM and NVM access across various sync levels. Additionally, 
it is evident that NVLog is the only solution that does not introduce any slowdown to the legacy disk FS.

\begin{figure}
    \centering
    \includegraphics[width=\linewidth]{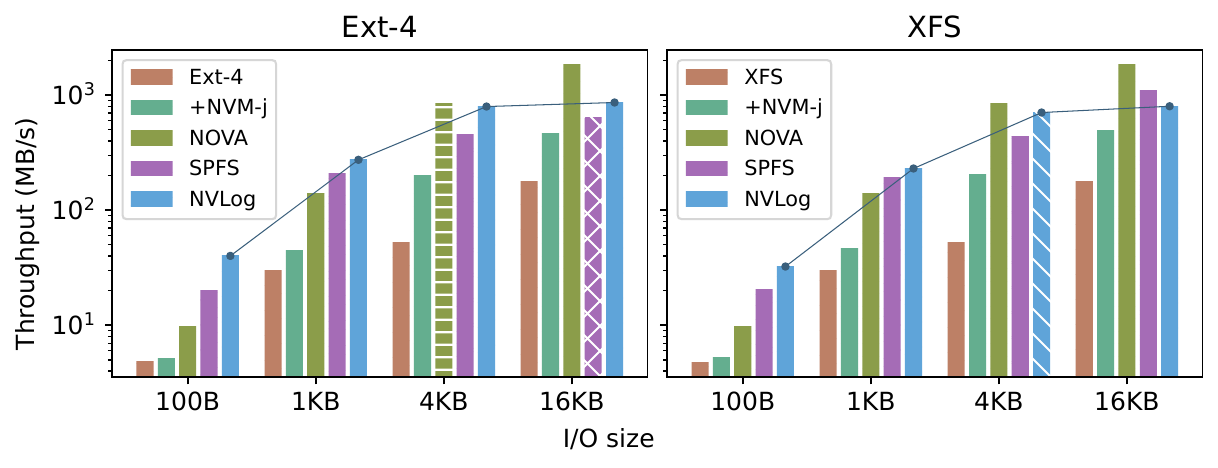}
    \caption{Sync performance under different I/O sizes. +NVM-j refers to the performance of placing the baseline file system's journal on NVM.}
    \label{fig:exp:sizes}
\end{figure}

\subsubsection{Pure Sync with Different I/O Sizes}

To demonstrate NVLog's performance characteristics under pure sync scenarios, we evaluated different systems with sequential sync writes across various I/O sizes. The results are shown in Figure \ref{fig:exp:sizes}. In these tests, NVLog consistently accelerates its underlying disk FS across all sizes, achieving up to a 15.09x acceleration with Ext-4 and 13.54x with XFS. We also evaluated the method of placing file system journal on NVM, marked as "+NVM-j" in the figure. Compared to the NVM-journaling method, NVLog can still provide up to 7.73x acceleration because (1) NVLog can bypass both the FS journaling and data write phases, whereas NVM-journaling only accelerates the journaling phase, and (2) NVLog is optimized for small-granularity writes and has a shorter call stack.

For small writes, NVLog can even outperform NOVA by up to 4.13x, thanks to its arbitrary-length log design that eliminates write amplification. However, NVLog fails to compete with NOVA (and SPFS in XFS) for large 16KB synchronous writes. The reasons are twofold: (1) NVLog cannot eliminate the software overhead of the underlying FS and page cache, and (2) NVLog writes to both DRAM and NVM, while NOVA and SPFS (after successful prediction) write only to NVM. The high performance of NOVA and SPFS in this scenario results from slowdowns in other scenarios, such as reads, as confirmed in our previous experiments. In line with our principles and goals, particularly \textbf{P1} and \textbf{P3}, we believe that NVLog’s moderate, non-aggressive acceleration is sufficient to enhance lower disk FS while providing better performance across a broad range of mixed daily scenarios.

\begin{figure}
    \centering
    \includegraphics[width=\linewidth]{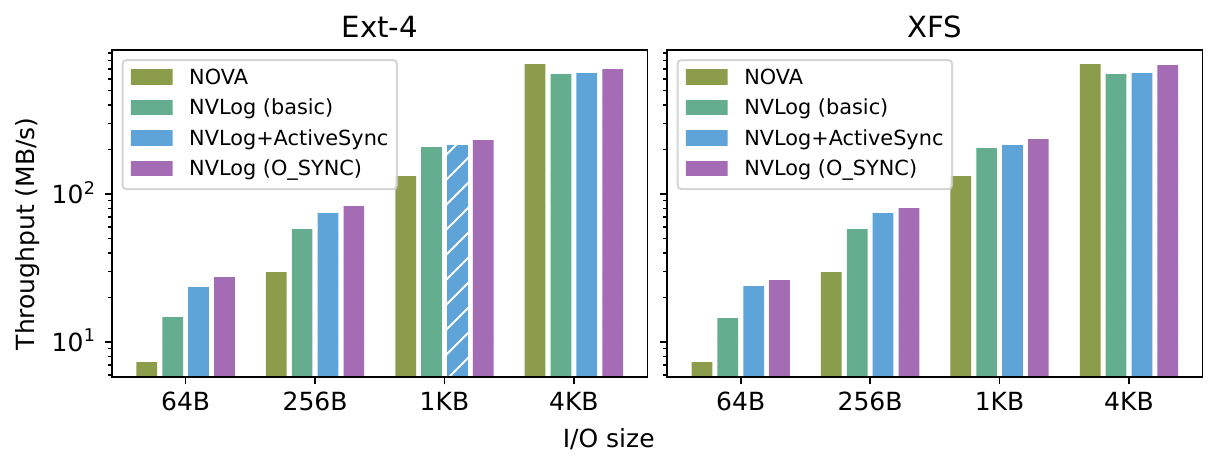}
    \caption{Active sync optimization.}
    \label{fig:exp:actsync}
\end{figure}

\subsubsection{Active Sync}

To investigate the performance benefits of the active sync mechanism, we test NVLog and NOVA with small sync writes of less than 4KB. In this test, synchronization is enforced by issuing \texttt{fsync} calls after each write operation. We also measured the performance of the \texttt{O\_SYNC} version NVLog as an upper bound. Figure \ref{fig:exp:actsync} shows the result. With active sync enabled, NVLog's performance is up to 1.62x faster than the vanilla one, and up to 3.22x faster than NOVA under 64B small writes. The smaller the I/O sizes, the more pronounced the effect of active sync, as it can reduce more write amplification. The active sync mechanism helps \texttt{fsync} achieve 86.21\% to 94.17\% of the performance of \texttt{O\_SYNC}, demonstrating the effectiveness of this optimization.

\subsubsection{Scalability}

\begin{figure}
    \centering
    \includegraphics[width=\linewidth]{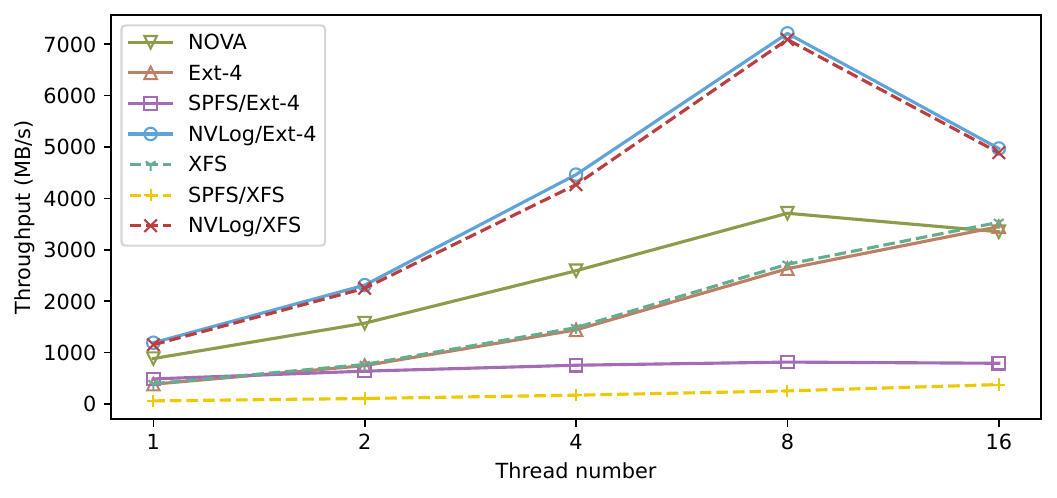}
    \caption{Scalability under random r/w test.}
    \label{fig:exp:scal}
\end{figure}

To measure the scalability of NVLog, we conduct a 4KB random read/write test with multiple threads accessing different files, varying the number of threads from 1 to 16. The read-write ratio is set to 1:1, with all writes being synchronized. The result is shown in Figure \ref{fig:exp:scal}. As the number of threads increases, NVLog scales well and outperforms all competitors. For Ext-4-based systems, NVLog outperforms NOVA, Ext-4, and SPFS by up to 1.94x, 3.11x, and 8.87x, respectively. For XFS-based systems, NVLog outperforms NOVA, XFS, and SPFS by up to 1.91x, 2.93x, and 28.18x, respectively. NVLog performs well because (1) it effectively uses DRAM and NVM separately to serve reads and writes, and (2) it does not introduce any additional locks to the software stack. SPFS shows poor performance due to its secondary index that fails to scale over multiple threads, and its read-after-sync slowdown.

Note that there is a performance degradation from 8 to 16 threads in both NOVA and NVLog. This is due to the saturation of the limited NVM write bandwidth, as our testbed only has two interleaved PM modules. The performance drop in NVLog is not related to our design, since NVLog does not introduce any additional locks. The test is stopped at 16 threads because our testbed only has 20 cores.

\begin{figure}
    \centering
    \includegraphics[width=\linewidth]{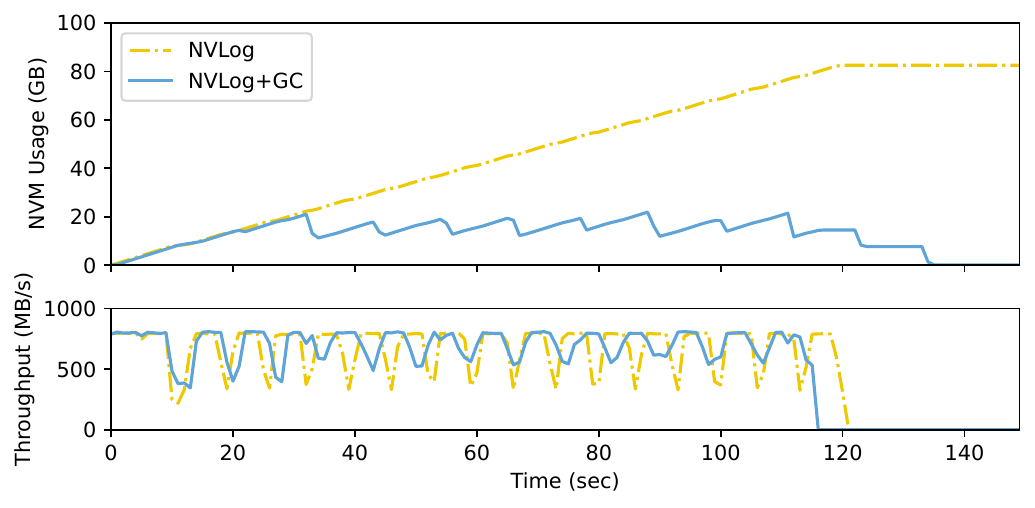}
    \caption{Garbage collection performance. The figure shows the NVM usage and the throughput of NVLog with or without garbage collection.}
    \label{fig:exp:gc}
\end{figure}

\begin{figure*}[t]
    \minipage{0.32\textwidth}
      \includegraphics[height=1.14in]{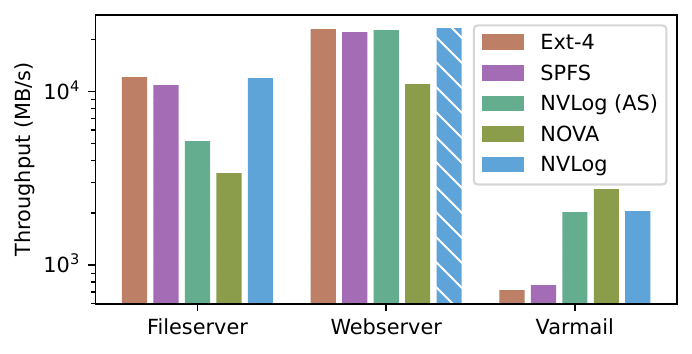}
      \caption{Filebench performance.}\label{fig:exp:filebench}
    \endminipage\hfill
    \minipage{0.32\textwidth}
      \includegraphics[height=1.14in]{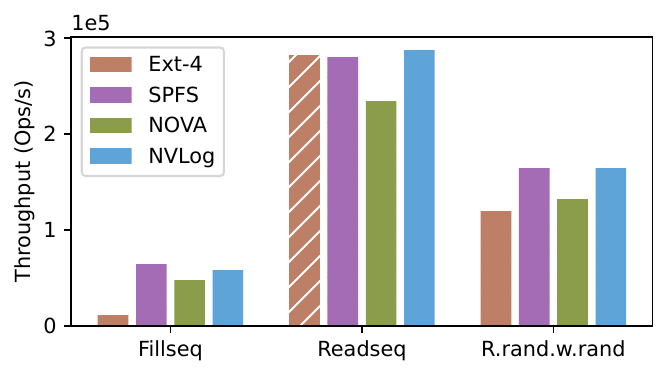}
      \caption{RocksDB performance.}\label{fig:exp:rocksdb}
    \endminipage\hfill
    \minipage{0.32\textwidth}%
      \includegraphics[height=1.14in]{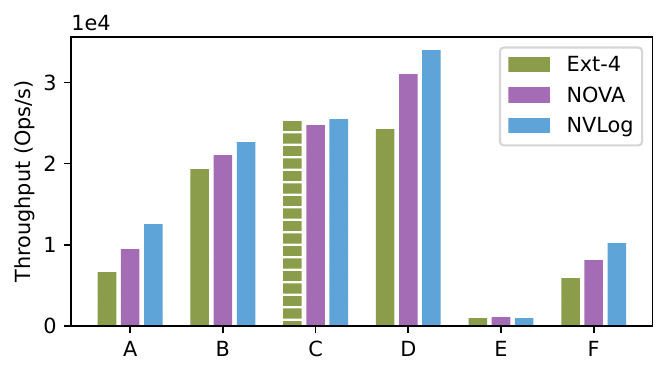}
      \caption{YCSB benchmark on SQLite.}\label{fig:exp:sqlite}
    \endminipage
\end{figure*}

\subsubsection{Garbage Collection}

We perform an 80GB sync write test on NVLog and track both the NVM usage and throughput to demonstrate the effectiveness and performance impact of garbage collection. The results are shown in Figure \ref{fig:exp:gc}. The garbage collection scan interval is set to 10 seconds, which is why the figure shows a drop in NVM usage every 10 seconds. In the first two rounds, the space reclamation is not obvious because the write-back process has not yet been activated. Throughout the entire process, the NVM usage of the garbage collection-enabled NVLog does not exceed 22GB and eventually drops to nearly zero. The performance fluctuations seen in the throughput are caused by the page allocator. Our implementation includes a per-CPU NVM page pool that helps improve performance. The observed performance drops indicate that the per-CPU pool is drained and is allocating new pages. Garbage collection alleviates this situation by returning unused pages back to the per-CPU pool. Overall, the garbage collector is effective and does not negatively impact performance.

\subsubsection{Capacity Limit}

We recognize that emerging NVM technologies may have relatively small capacities, so we test the performance of NVLog under capacity-limited conditions. We restrict the available NVM size to 10GB (about half of the peak usage reported in Figure \ref{fig:exp:gc}) and then use db\_bench (will be introduced in Section \ref{sec:exp:rocksdb}) with its typical workloads to test NVLog's performance. For read (\texttt{readseq}) and uniformly random read-write (\texttt{readrandomwriterandom}) workloads, limiting the NVM capacity has no impact on performance. For the fully synchronous write (\texttt{fillseq}) workload, performance drops by 57\% compared to the unlimited case shown in Figure \ref{fig:exp:rocksdb}, but it is still 2.25x faster than Ext-4.

\subsection{Macrobenchmarks}

Due to the similar performance trends observed across different disk FS baselines in the microbenchmarks, we will use only Ext-4 as the baseline in the macrobenchmarks.

\subsubsection{Filebench}

Filebench \cite{Filebench} provides 3 representative macrobenchmark scripts to simulate different server workloads: \texttt{fileserver}, a non-sync, write-intensive workload with a 1:2 r/w ratio; \texttt{webserver}, a read-intensive workload with a 10:1 r/w ratio; and \texttt{varmail}, a balanced read/sync-write 1:1 workload with small I/O size. The detailed configurations of these workloads are listed in Table \ref{tab:filebench_config}. 

\begin{table}[h!]
    \centering
    \caption{Filebench workload configurations.}
    \resizebox{\linewidth}{!}{
    \begin{tabular}{c||c|c|c|c|c}
        \toprule
        \makecell{Workload} & \makecell{File size \\ (avg)} & \makecell{I/O size \\ (r/w)} & \makecell{Threads} & \makecell{R/W \\ ratio} & \makecell{\# of files}  \\
        \hline 
        Fileserver & 128KB & 1MB/16KB & 16 & 1:2 & 10000 \\
        Webserver & 64KB & 1MB/16KB & 16 & 10:1 & 1000 \\
        Varmail & 16KB & 1MB/16KB & 16 & 1:1 & 10000 \\
        \bottomrule

    \end{tabular}
    }
    
    \label{tab:filebench_config}
\end{table}

The result of the test is shown in Figure \ref{fig:exp:filebench}. In the \texttt{fileserver} and \texttt{webserver} workloads, NVLog, SPFS, and Ext-4 exhibit similar performance, significantly outperforming NOVA due to their use of a fast DRAM page cache. For example, NVLog demonstrates 3.55x and 2.10x the performance of NOVA in the \texttt{fileserver} and \texttt{webserver} workloads, respectively. 
NVLog (AS) requires all writes to be synchronized, which results in worse performance than standard NVLog in \texttt{fileserver}. Nonetheless, NVLog (AS) still outperforms NOVA by 1.53x, primarily due to the DRAM-enabled reads.
In the \texttt{varmail} workload, NVLog is 2.84x and 2.65x faster than Ext-4 and SPFS, respectively, but 25.98\% slower than NOVA. SPFS fails to effectively accelerate Ext-4 in \texttt{varmail} because it requires a prediction period before absorbing sync writes to NVM. However, \texttt{varmail} synchronously writes to scattered files only twice per file, preventing SPFS from predicting and absorbing most of these scattered sync writes. The relatively lower performance of NVLog compared to NOVA in \texttt{varmail} is due to the double-write to both DRAM and NVM. Nevertheless, we believe that retaining the DRAM cache helps to provide a more balanced performance in daily workloads, where sync operations constitute only a small portion.

\subsubsection{RocksDB}
\label{sec:exp:rocksdb}

RocksDB \cite{RocksDB} is an LSM-tree-based key-value database for server workloads. Data written to RocksDB is first recorded in a write-ahead log (WAL) and then asynchronously written to the LSM tree (SST files). Reading data from RocksDB involves reading from the SST files. RocksDB includes db\_bench as its testing suite. To demonstrate NVLog's performance under different conditions, we selected three tests in db\_bench: sequential write (\texttt{fillseq}), sequential read (\texttt{readseq}), and multi-threaded mixed read/write (\texttt{readrandomwriterandom}). The value size is set to 4KB, and the level 1 file size is set to 512MB. We initially run a \texttt{fillseq} to create the database, enable sync mode for each test, and remove the database after each test. Figure \ref{fig:exp:rocksdb} shows the performance of RocksDB.

For \texttt{fillseq}, SPFS, NOVA, and NVLog are 5.83x, 4.33x, and 5.23x faster than Ext-4, respectively. The low performance of Ext-4 is due to its WAL sync writes suffering from the low speed of the disk. NOVA performs slower than SPFS and NVLog because of its write amplification for small metadata writes, which is a result of its copy-on-write design.

For \texttt{readseq}, Ext-4 and NVLog perform similarly, and both outperform NOVA. This is because, in NVLog and Ext-4, read operations are served by the fast-path DRAM, whereas NOVA can only perform reads on NVM. SPFS also provides comparable high speed, as it avoids its inherent read-after-write slowdown. Specifically, RocksDB reads from SST files, which are previously written to disk in large chunks (tens or hundreds of MB) synchronously. SPFS does not absorb sync writes larger than 4MB, allowing it to continue serving RocksDB reads from DRAM rather than NVM.

For \texttt{readrandomwriterandom}, NVLog performs 1.38x faster than Ext-4 and 1.24x faster than NOVA. The advantage of NVLog stems from its cooperative DRAM-NVM design. SPFS achieves similar performance to NVLog, once again due to its ability to skip large bulk syncs.

Overall, NVLog and SPFS both achieve balanced performance on RocksDB, with higher speeds than Ext-4 for writes and comparable speeds for reads, while NOVA provides reasonable write speeds but deteriorated read speeds. However, note that SPFS encountered several crashes when attempting to clean up the existing database and failed to run under RocksDB's \texttt{O\_DIRECT} mode. By contrast, NVLog and NOVA both demonstrated good robustness during the test.

\subsubsection{SQLite}

SQLite \cite{SQLiteHomePage} is a lightweight embedded SQL database engine widely used across PC and mobile applications. To evaluate whether SQLite can benefit from NVLog, we use YCSB \cite{cooperBrianfrankcooperYCSB2024, leeLs4154YCSBcpp2024}, a common benchmark for database performance. YCSB includes six different workloads, labeled \texttt{A} through \texttt{F}, to cover a variety of application scenarios. In this test, SQLite is in FULL synchronous mode, and the record size is 4KB. 
The user space cache is set to 0 to fully demonstrate the performance of the underlying software stack. 
The result of SQLite’s performance is illustrated in Figure \ref{fig:exp:sqlite}.

In workloads \texttt{A}, \texttt{B}, \texttt{D}, and \texttt{F}, which involve writing to the database, NVLog outperforms both Ext4 and NOVA. NVLog accelerates Ext-4 by up to 1.91x due to the high persistence speed of NVM. NVLog also surpasses NOVA by up to 1.33x because its arbitrary-length log structure and active sync mechanism efficiently handle small metadata updates. In workloads \texttt{C} and \texttt{E}, which are read-only, the performance differences among the tested systems are minimal. 
Since we have eliminated the impact of user cache, this is likely because the query execution time dominates, making the performance differences in page cache and storage media less significant. 
SPFS did not appear in this experiment due to recurring crashes during testing.

\section{Conclusion}
\label{sec:conclusion}

In this paper, we propose 
NVLog, an NVM-based write-ahead log designed to transparently accelerate traditional disk file systems. 
NVLog is the only approach that effectively balances performance across DRAM, NVM, and disk, ensuring that the enhancement does not introduce any slowdowns to the existing storage stack. Our experiments demonstrate that NVLog outperforms previous solutions in most common use cases, while also maintaining a stable and lightweight characteristic.

\section*{Acknowledgments}

We would like to thank our shepherd, Peter Macko, for his detailed suggestions and guidance during the revision phase of our paper. We also wish to express our gratitude to Professor Guangyan Zhang from Tsinghua University and Professor Cheng Li from University of Science and Technology of China for their guidance and assistance. The corresponding author of this paper is Juncheng Hu.

This work is supported by the National Key R\&D Program under Grant No. 2024YFB3310200, and by the Key Scientific and Technological R\&D Plan of Jilin Province of China under Grant No. 20230201066GX, and by the Central University Basic Scientific Research Fund Grant No. 2023-JCXK-04.

\section*{Availability}


The prototype of NVLog is implemented and available at  \url{https://github.com/BugJLU/NVLog}.

\bibliographystyle{plain}

\appendix
\section{Artifact Appendix}

\subsection*{Abstract}

The artifact provides the source code of our NVLog prototype. It includes a modified Linux kernel that implements NVLog, and a series of utilities for configuration and evaluation. 

\subsection*{Scope}

This artifact includes a Linux kernel with the implementation of designs described in the paper, along with relevant scripts and documentation. It supports performance evaluation and facilitates a deeper understanding of the paper's design details.

\subsection*{Contents}

The NVLog artifact consists of three main components:
\begin{enumerate}
    \item \textbf{Modified Linux Kernel (\texttt{/linux-5.15.125}):} This is the core of the artifact, containing the implementation of NVLog's design.
    \item \textbf{Utility Scripts:} These include scripts for setting up the environment and compiling the kernel (\texttt{/dev\_vm}), as well as scripts for configuring NVLog in user space (\texttt{/utils}).
    \item \textbf{Documentation:} This includes \texttt{/README.md} and additional README files linked within.
\end{enumerate}
Additionally, the artifact provides instructions and scripts for performance evaluation of NVLog in \texttt{/ae}, supporting artifact evaluation and future work.

\subsection*{Hosting}

The artifact is available on GitHub: \url{https://github.com/BugJLU/NVLog}.

\subsection*{Requirements}

To run this prototype and reproduce the experimental results in this paper, Intel Optane PMEM is required. 

\subsection*{Evaluation}

Due to the large amount of experimental data presented in the paper and the slow performance of standard Ext-4 in synchronization test cases, reproducing all experiments during artifact evaluation could incur significant time cost. To address this, we have streamlined the experiments into three main claims and provided the necessary experimental scripts to validate these claims. The following are the three claims:

\begin{description}    
    \item[C1:] \textbf{NVLog can leverage NVM and DRAM to separately handle synchronous writes and other requests, achieving optimal performance under complex mixed read, asynchronous write, and synchronous write workloads.} To validate this claim, we designed performance tests with different read/write ratios (R/W=0/10, 3/7, 5/5, 7/3), where 50\% of the writes are set as synchronous. NVLog is expected to outperform NOVA, SPFS, and Ext4 in these tests.
    \item[C2:] \textbf{When performing sub-page granularity synchronous writes, NVLog can achieve optimal performance by fully utilizing the byte-granularity characteristics of NVM.} To validate this claim, we conducted synchronous write performance tests at a 64B granularity. NVLog is expected to outperform NOVA, SPFS, and Ext4 in these tests.
    \item[C3:] \textbf{Due to NVLog's design, which supports garbage collection, it temporarily occupies only a small amount of NVM.} To validate this claim, we conducted an 80GB synchronous write test on NVLog. During most of the process, the NVM usage should be less than the write volume, and after garbage collection is completed, the NVM usage should be less than 1\% of the total write volume.
\end{description}



\end{document}